\documentstyle[preprint,aps]{revtex} 
\tightenlines 
\begin{document}

\draft
\title{Minimal Length Uncertainty Relation and Hydrogen Atom}
\author{F. Brau\thanks{E-mail: 
fabian.brau@umh.ac.be}} 
\address{Universit\'{e} de Mons-Hainaut,
B-7000 Mons, BELGIQUE}
\date{\today}

\maketitle

\begin{abstract}

We propose a new approach to calculate perturbatively the effects of a
particular
deformed Heisenberg algebra on energy spectrum. We use this method to
calculate the harmonic oscillator spectrum and find that corrections are in
agreement with a previous calculation. Then, we apply this approach to
obtain the hydrogen atom spectrum and we find that splittings of degenerate
energy levels appear. Comparison with experimental data yields an interesting
upper bound for the deformation parameter of the Heisenberg algebra.
\end{abstract}

\section{Introduction}
\label{sec:intro}

Study of modified Heisenberg algebra, by adding certain small corrections to
the
canonical commutation
relations, arouse a great interest for some years 
(see for example~\cite{kemp95,hinr96,kemp97b,kemp97,brou99}). These
modifications yield  
new short distance structure characterized by a finite minimal uncertainty
$\Delta x_0$ in 
position measurement. The existence of this minimal observable length has been
suggested by 
quantum gravity and string theory~\cite{gros88,amat89,magg93,amel97,haro98}.
In this
context, the new short
distance behavior would arise at 
the Planck scale, and $\Delta x_0$ would correspond to a fundamental
quantity closely linked with the structure of the space-time \cite{kemp98}.
This feature constitutes a part of
the motivation to study 
the effects of this modified algebra on various observables. 

Recently, it has been suggested that this formalism could 
be also used to describe, as an effective theory, non-pointlike particles,
e.g. hadrons, 
quasi-particles or collective excitations~\cite{kemp97b}. In this case,
$\Delta x_0$ is interpreted as 
a parameter linked with the structure of particles and their finite size.
In the work~\cite{kemp97b} the $d$-dimensional isotropic harmonic oscillator
was solved, in the context of a nonvanishing $\Delta x_0$, with 
a special interest to the 3-dimensional case. This calculation shows that
splittings of 
usual degenerate energy levels appear, leaving only the degeneracy due to
the independence 
of the energy on the azimuthal quantum number, $m$. It has been also indicated
that application 
to the hydrogen atom should yield the relation between the scale of a
non-pointlikeness of 
the electron and the scale of the caused effects on the hydrogen spectrum.
Indeed, the high 
precision of the experimental data for the transition $1S-2S$ \cite{udem97},
for example, can yield an 
interesting upper bound for the possible, in the sense studied here, finite
size of the electron. 

The purpose of this work is to continue to investigate whether the Ansatz
concerning the deformation of the Heisenberg algebra, with suitably adjusted
scale, may also serve for an effective low energy description of non-pointlike
particles. In this way, we calculate corrections to the hydrogen
spectrum using the minimally modified Heisenberg algebra, i.e. which preserves
the
commutation relations between position operators. To perform this calculation
we propose a new approach which allow to solve the Schr\"{o}dinger equation in
the position representation. This method leads to the correct harmonic
oscillator spectrum found in Ref.~\cite{kemp97b}. Application to hydrogen atom
shows that splittings of the usual degenerate energy levels are also present
and
that these corrections cannot be seen experimentally if $\Delta x_0$ is
smaller than $0.01$ fm.

\section{Method}
\label{sec:theory}

The modified Heisenberg algebra studied here, as it has been done in
Ref.~\cite{kemp97b}, is defined by the following commutation relations ($\hbar
= c = 1$)
\begin{eqnarray}
\label{eq1}
\nonumber
\left[\hat{X}_i,\hat{P}_j\right] &=& i\left(\delta_{ij}+\beta \delta_{ij}
\hat{P}^2
+\beta' \hat{P}_i \hat{P}_j\right), \\
\left[\hat{P}_i,\hat{P}_j\right] &=& 0,
\end{eqnarray}
where $\hat{P}^2 = \sum_{i=1}^{3} \hat{P}_i \hat{P}_i$ and where $\beta,
\beta' > 0$ are considered as small quantities of the first order. In this
paper, we study only
the case $\beta'=2\beta$, which leaves the commutation relations between the
operators $\hat{X}_i$ unchanged~\cite{kemp97b}, i.e.
$\left[\hat{X}_i,\hat{X}_j\right] = 0$.
This constitutes the minimal extension of the Heisenberg algebra and is thus
of
a special interest.

To calculate a spectrum for a given potential, we must find a representation
of the operators $\hat{X}_i$ and $\hat{P}_i$, involving position variables
$x_i$ and partial derivatives with respect to these position variables, which
satisfies Eqs.~(\ref{eq1}), and solve the corresponding Schr\"{o}dinger
equation:
\begin{equation}
\label{eq2}
\left[ \frac{\hat{P}^2}{2m} + V\left(\vec{\hat{X}}\right)\right]\,
\Psi(\vec{x}\,) = E\, \Psi(\vec{x}\,).
\end{equation}
It is straightforward to verify that the following representation fulfill the
relations (\ref{eq1}), in the first order in $\beta$,
\begin{eqnarray}
\label{eq3}
\nonumber
\hat{X}_i\ \Psi(\vec{x}\,) &=& x_i \Psi(\vec{x}\,), \\
\hat{P}_i\, \Psi(\vec{x}\,) &=& p_i \left(1+\beta \vec{p}\,^2\right)
\Psi(\vec{x}\,) \quad \rm{with} \quad p_i=\frac{1}{i}
\frac{\partial}{\partial x_i}.
\end{eqnarray}
Neglecting terms of order $\beta^2$, the Schr\"{o}dinger equation (\ref{eq2})
takes the form
\begin{equation}
\label{eq4}
\left[ \frac{\vec{p}\,^2}{2m} +\frac{\beta}{m} \vec{p}\,^4 +
V(\vec{x}\,)\right]\,
\Psi(\vec{x}\,) = E\, \Psi(\vec{x}\,).
\end{equation}
This is the ordinary Schr\"{o}dinger equation with an additional term
proportional to $\vec{p}\,^4$. As this correction is assumed to be small, we
calculate its effects on energy spectra in the first order of perturbations.
The evaluation of the spectrum to the first order in the deformation
parameter $\beta$ leads to
\begin{equation}
\label{eq5}
E_{k}=E_{k}^0+\Delta E_{k},
\end{equation}
where $k$ denotes the set of quantum numbers which labels the energy level,
and
where $\Delta E_{k}$ are the eigenvalues of the matrix
\begin{equation}
\label{eq6}
\frac{\beta}{m} \langle\Psi_{k}^0(\vec{x}\,)|\,\vec{p}\,^4\,
|\Psi_{k'}^0(\vec{x}\,)
\rangle \equiv
\frac{\beta}{m} \langle k| \vec{p}\,^4 |k'\rangle,
\end{equation}
where $\Psi_{k}^0(\vec{x}\,)$ are solutions of (\ref{eq4}) with
$\beta=0$. This matrix is computed with all the wave functions corresponding
to the unperturbed energy level $E_{k}^0$.
This is a $g \times g$ matrix where $g$ is the multiplicity of the
state $E_{k}^0$ considered. In general, $\Delta E_{k}$ takes $f$
($f \leq g$) different values which removes the degeneracy of some energy
levels. For
an arbitrary interaction $V(\vec{x}\,)$ used in the Schr\"{o}dinger equation, 
the matrix (\ref{eq6}) is non-diagonal. But, since we know the action of
$\vec{p}\,^2$ (from Eq.~(\ref{eq4})) on the unperturbed wave functions, the
expression of the
matrix elements, for a central potential, can be written as
\begin{equation}
\label{eq6b}
4\beta m \left(\left(E^0_{n,\ell}\right)^2 \delta_{n
n'}-(E_{n,\ell}^0+E_{n',\ell}^0)\langle
n\ell m| V(r) |n'\ell m\rangle +\langle n\ell m| V(r)^2 |n'\ell m\rangle
\right)\, \delta_{\ell \ell'} \delta_{m m'},
\end{equation}
and, in the cases studied here, there are no degenerate states with equal
values of the angular momentum $\ell$ and the azimuthal quantum number $m$
which have different values of radial quantum number $n$. Thus the matrix
(\ref{eq6}) is diagonal and the correction to the spectrum can be written a
\begin{equation}
\label{eq7}
\Delta E_{n,\ell}= 4\beta m \left(\left(E_{n,\ell}^0\right)^2-2E_{n,\ell}^0\,
\langle n\ell m|
V(r)|n\ell m
\rangle+\langle n\ell m|V(r)^2|n\ell m \rangle\right).
\end{equation}
This nice relation can be simplified if one considers power-law central
potential, $V(r)\sim r^p$. In this case, the virial theorem gives
\begin{equation}
\label{eq8}
\langle n\ell m|  V(r)|n\ell m \rangle= \frac{2}{p+2}\, E_{n,\ell}^0,
\end{equation}
which leads to the following form for the expression of the energy level shift
in the first order in $\beta$:
\begin{equation}
\label{eq9}
\Delta E_{n,\ell}= 4\beta m \left(\left(E_{n,\ell}^0\right)^2
\left(\frac{p-2}{p+2}\right)+\langle
n \ell m|V(r)^2|n\ell m \rangle\right).
\end{equation}
This simple expression will allow us to find the correction of the harmonic
oscillator and hydrogen spectra just by calculating the mean value of the
square of the potential.

\section{Harmonic Oscillator}
\label{subsec:firstf}

For this potential, the energy level shift is only given by the mean value of
the square of the potential. The normalized unperturbed wave function of the
harmonic oscillator reads
\begin{equation}
\label{eq10}
\Psi^0_{n \ell m}(\vec{r}\,)= \lambda^{3/2} \sqrt{\frac{2\
n!}{\Gamma(n+\ell+3/2)}}\ (\lambda r)^\ell\, e^{-(\lambda r)^2/2}\,
L^{\ell+1/2}_{n}\left((\lambda r)^2\right)\, Y_{lm}(\theta,\varphi),
\end{equation}
where $\lambda=\sqrt{m\omega}$ and $L^{\alpha}_n(x)$ are Laguerre
polynomials \cite [p. 1037]{grad80}. $n$ is the radial quantum number. Using
the change of variable $x=(\lambda
r)^2$, the energy shift is found to be:
\begin{equation}
\label{eq11}
\Delta E_{n,\ell}= \frac{4\beta m (n!) k^2}{\lambda^4 \Gamma(n+\ell+3/2)} \
\int_0^{\infty} x^{\ell+5/2}\, e^{-x}\,
\left[L^{\ell+1/2}_{n}(x)\right]^2\,dx, 
\end{equation}
where $2k=m\omega^2$ is the strength of the oscillator force. The calculation
of the remaining integral is straightforward. Knowing the
following relations concerning the Laguerre polynomials \cite [p. 1037, p.
844]{grad80}
\begin{eqnarray}
\label{eq12}
L^{\alpha-1}_{n}(x)&=&L^{\alpha}_{n}(x)-L^{\alpha}_{n-1}(x),\\
\label{eq13}
\int_0^{\infty}  e^{-x}\,
x^{\alpha}\,L^{\alpha}_{n}(x)\,L^{\alpha}_{m}(x)\, dx &=&
\frac{\Gamma(\alpha+n+1)}{n!}\, \delta_{nm},
\end{eqnarray}
we obtain the expression of the harmonic oscillator spectrum for the modified
Heisenberg algebra (\ref{eq1}):
\begin{equation}
\label{eq14}
E_{n,\ell}= \omega (2n+\ell+3/2)+ (\Delta x_0)^2 \frac{m\omega^2}{5}
(6n^2+9n+6nl+\ell^2+4\ell+15/4), 
\end{equation}
where $\Delta x_0=\sqrt{5\beta}$. This formula reproduces exactly the
splittings calculated in Ref.~\cite{kemp97b} using another approach. 
Because the dependence on quantum numbers of the correction term is not of the
form $f(2n+\ell)$, we obtain splittings of degenerate levels. But the energy
does not depend on the azimuthal quantum number $m$ and each level remains
$(2\ell+1)$-fold degenerate.

This example shows the usefulness of this approach which provides, with simple
calculations, an analytical expression of the energy shift. The main interest
of this method is that it can easily be used to solve analytically or
numerically
other problems, such as the Coulomb problem which is solved in the next
section. 

\section{Hydrogen Atom}
\label{subsec:secondf}
As we mentioned in the Introduction, the evaluation of corrections of the
energy spectrum can provide information concerning, in the sense studied here,
an
assumed finite size of electrons. The method used here to describe
non-pointlike particles neglects the internal structure degree of freedom. But
obviously these effects have much smaller order of magnitude and thus can be
omitted. 

The normalized unperturbed wave function of the hydrogen atom reads
\begin{equation}
\label{eq15}
\Psi^0_{n \ell m}(\vec{r}\,)= (2\gamma_n)^{3/2}
\sqrt{\frac{(n-\ell-1)!}{2n(n+\ell)!}}\ (2\gamma_n r)^\ell\, e^{-\gamma_n r}\,
L^{2\ell+1}_{n-\ell-1}(2\gamma_n r)\, Y_{lm}(\theta,\varphi),
\end{equation}
where $\gamma_n=m\alpha/n$ and $\alpha$ is the fine structure constant. $n$ is
the principal quantum number and $\ell$ varies between $0$ and $n-1$. The
change of variable $x=2\gamma_n r$ allow to write the energy shift as
\begin{equation}
\label{eq16}
\Delta E_{n,\ell}= -12\beta m \left(E^0_{n,\ell}\right)^2+8\beta m \gamma_n^2
\alpha^2\,
\frac{(n-\ell-1)!}{n(n+\ell)!} \
\int_0^{\infty} x^{2\ell}\, e^{-x}\,
\left[L^{2\ell+1}_{n-\ell-1}(x)\right]^2\,dx. 
\end{equation}
Like for the harmonic oscillator problem, the evaluation of this integral is
quite simple. Indeed, using the following relation for Laguerre polynomials
\cite [p.
1038]{grad80}
\begin{equation}
\label{eq17}
\sum_{m=0}^{n} L_m^{\alpha}(x) =L_n^{\alpha+1} (x),
\end{equation}
with the relation (\ref{eq13}) and the following summation formula
\begin{equation}
\label{eq18}
\sum_{p=0}^{b} \frac{(p+a)!}{p!} =\frac{(a+b+1)!}{(1+a) b!},
\end{equation}
the expression of the hydrogen spectrum, in the first order in the deformation
parameter $\beta$, reads
\begin{equation}
\label{eq19}
E_{n,\ell}=-\frac{m\alpha^2}{2n^2}+\left(\Delta x_0\right)^2 \, \frac{m^3
\alpha^4}{5}\, 
\frac{(4n-3(\ell+1/2))}{n^4 (\ell+1/2)}.
\end{equation}
This formula shows that the corrections to the spectrum are always positive.
The value of this additional term is maximum for the ground state and for
each value of $n$, the maximal contribution is obtained for $\ell=0$ levels.
Like in the harmonic oscillator case, the correction term, which depends
explicitly on $\ell$, lifts the degeneracy of energy levels which
remain, however, $(2\ell+1)$-fold degenerate.

The accuracy concerning the measurement of the frequency of the radiation
emitted
during the transition $1S-2S$ is about 1 kHz \cite{udem97}.
Thus the energy difference between this two levels is determined with a
precision about $10^{-12}$ eV. Then, if we suppose that effects of
finite size of electrons cannot yet be seen experimentally, we find 
\begin{equation}
\label{eq20}
\Delta x_0 \leq 0.01\ \rm{fm}.
\end{equation}
But corrections calculated here could already play a role in the theoretical
description of the hydrogen atom since the accuracy of theoretical
calculations is less good than the precision of experimental data. The main
theoretical error is the determination of the proton charge radius. Thus, at
this moment, confrontation between experimental data and standard theoretical
calculations cannot exclude the effects studied in this paper.

Nevertheless, the upper bound (\ref{eq20}) seems to be reasonable. Moreover, a
naive argument can give an order of magnitude of an ``experimental" upper
bound
for the finite size of the electron. Indeed, a lower bound for the mass of an
excited state of the electron is about 85 GeV \cite{pdg98}. Thus a photon with
an energy of about 85 GeV cannot excited an electron. In a first
approximation, this means that the resolution obtained with this photon is not
sufficient to detect a finite size of electrons. The wave length of a such
photon could constitute an upper bound for the size of electrons,
\begin{equation}
\label{eq21}
\Delta x_0 \leq \lambda \sim 0.015\ \rm{fm}.
\end{equation}
This naive argument applied to the nucleon and its first radial excitation
N(1440) yields a size of about 2.5 fm which is the correct order of magnitude.

Thus in a (very?) near future, with improvement of the accuracy of
experimental data and above all improvement of the precision of standard
theoretical
calculations, it could be possible either to lower down the upper bound
(\ref{eq20}) or detect the existence of a non-vanishing $\Delta x_0$.

\section{Summary}
\label{sec:summary}

We have proposed a new formulation of the Schr\"{o}dinger equation which takes
into account the deformation of the Heisenberg algebra in the first order in
the deformation parameter $\beta$. This modified algebra introduces a minimal
observable length in the uncertainty relations. It has been proposed in
Ref.~\cite{kemp97b} that this framework could be used to describe
non-pointlike
particles as an effective low energy theory, neglecting their internal
structure
degree of freedom. The minimal length $\Delta x_0$ would be then linked with
the non-pointlikeness of particles.

In Sec.~\ref{subsec:firstf}, we have calculated, with the new approach, the
corrections to the
harmonic oscillator spectrum which are in agreement with those derived in
a previous calculation using another approach~\cite{kemp97b}. Note that
this method can be generalized to other dimensions. In particular, we have
verify that the spectrum of the 1-dimensional harmonic oscillator is in
agreement with that found in Ref.~\cite{kemp95}. Moreover, the wave function
in the position space can also be calculated, in the first order in $\beta$,
just as various observables associated to the systems studied. 

In Sec.~\ref{subsec:secondf}, we have used this method to obtained the
corrections to the hydrogen atom spectrum. Comparison with the experimental
data for the transition $1S-2S$ \cite{udem97} yields a plausible upper bound
for the non-pointlikeness $\Delta x_0$ of the electron which is about $0.01$
fm.

The formulation of the Schr\"{o}dinger equation proposed here could prove to
be very useful to study properties of some systems and their various
associated observables in the context of the deformed Heisenberg algebra
studied here.

\acknowledgments

We thank Professor F. Michel for stimulating discussions, and Professor Y.
Brihaye for reading the manuscript. We would like to thank IISN for financial
support.

\end{document}